\pdfoutput=1
\documentclass[aps,prl,nofootinbib,twocolumn,superscriptaddress,showpacs,floatfix]{revtex4-1}

\usepackage{bm}
\usepackage{amsfonts}
\usepackage[pdftex]{graphicx}
\usepackage[usenames,dvipsnames]{color}
\usepackage{amssymb}
\usepackage{amsmath}
\usepackage{bbold}

\usepackage{bbm}
\usepackage{amsthm}
\usepackage{hyperref}
\hypersetup{
        colorlinks=false,
}
\usepackage{times}

\usepackage{ulem}
\usepackage[english]{babel}
\usepackage{blindtext}
\usepackage{subfigure}

\newcommand{\ket}[1]{\left| #1 \right>}

\newcommand{\braket}[2]{\left< #1 | #2 \right>}

\newcommand{\fcite}[1]{\ensuremath{^{[{\color{red}x}]}}}
\let\oldmarginpar\marginpar
\renewcommand\marginpar[1]{\-\oldmarginpar[\raggedleft\tiny\color{red} #1]%
{\raggedright\tiny #1}}

\DeclareMathAlphabet\mathbfcal{OMS}{cmsy}{b}{n}
\graphicspath{{Figures/}}

\begin{document}

\title{Machine learning quantum phases of matter beyond the fermion sign problem}
\date{\today}

\author{Peter Broecker}
\affiliation{Institute for Theoretical Physics, University of Cologne, 50937 Cologne, Germany}
\author{Juan Carrasquilla}
\affiliation{Perimeter Institute for Theoretical Physics, Waterloo, Ontario N2L 2Y5, Canada}
\author{Roger G. Melko}
\affiliation{Perimeter Institute for Theoretical Physics, Waterloo, Ontario N2L 2Y5, Canada}
\affiliation{Department of Physics and Astronomy, University of Waterloo, Ontario, N2L 3G1, Canada}
\author{Simon Trebst}
\affiliation{Institute for Theoretical Physics, University of Cologne, 50937 Cologne, Germany}


\begin{abstract}
State-of-the-art machine learning techniques promise to become a powerful tool in statistical mechanics via their capacity to distinguish different phases of matter in an automated way. Here we demonstrate that convolutional neural networks (CNN) can be optimized for quantum many-fermion systems such that they correctly identify and locate quantum phase transitions in such systems. Using auxiliary-field quantum Monte Carlo (QMC) simulations to sample the many-fermion system, we show that the Green's function  (but not the auxiliary field) holds sufficient information to allow for the distinction of different fermionic phases via a CNN. 
We demonstrate that this QMC + machine learning approach works even for  systems exhibiting a severe fermion sign problem where conventional approaches to extract information from the Green's function, e.g.~in the form of equal-time correlation functions, fail. We expect that this capacity of hierarchical machine learning techniques to circumvent the fermion sign problem will drive novel insights into some of the most fundamental problems in statistical physics.
\end{abstract}

\maketitle


In quantum statistical physics, the sign problem refers to the generic inability of quantum Monte Carlo (QMC) approaches to tackle fermionic systems
with the same unparalleled efficiency it exhibits for unfrustrated bosonic systems.
At its most basic level, it tracks back to the expansion of the partition function 
of a quantum mechanical system in terms of $d+1$ dimensional classical configurations that have both positive and negative (or complex) statistical weights, 
thus invalidating their usual interpretation as a probability distribution \cite{Hirsch1982,Loh1990}.  
In some specific cases, canonical transformations
or basis rotations are known that completely eliminate the negative weights \cite{MeronCluster,FermionBag,MajoranaBasis}, resulting in sign-problem free models, sometimes called ``stoquastic'' \cite{Stoquastic} or ``designer'' \cite{Designer} Hamiltonians.  However, the lack of a general systematic procedure for such transformations \cite{TroyerWiese2005} preclude many, if not most, quantum
Hamiltonians from being simulated with unbiased QMC methods.  
This includes one of the most fundamental problems in statistical physics -- the many-electron system, which is known to give rise to 
some of the most intriguing collective phenomena such  as the formation of high-temperature superconductors \cite{Superconductors}, non-Fermi liquids \cite{Schofield1999,Rosch2007}, 
or Mott insulators with fractionalized excitations \cite{Balents2010}.

When tackling sign-problematic Hamiltonians with QMC approaches
a common procedure \cite{LandauBinder} consists of two steps: (1) taking the absolute value of the configuration weight, thereby allowing interpretation as a 
probability amenable to sampling; and (2) precisely compensating for this by weighing observables (such as two-point correlation functions) with the excluded sign.  
While this procedure allows, in principle, for an unbiased evaluation of observables, it introduces changes into 
the sampling scheme in two distinct ways.  First, the exclusion of the sign in step (1) affects the region of configuration space that is effectively sampled. 
To what extent this modified sampling imposes severe restrictions or rather subtle constraints very much depends on the actual QMC flavor, such as 
world-line versus auxiliary-field approaches.
Second, this modified sampling necessitates the sign reweighing of step (2) in any proper statistical analysis. It is, however, precisely this step where
the sign problem ultimately manifests itself in a statistical variance of estimators that grows {\it exponentially} in system size and inverse temperature.

In this paper, we examine an approach by which these two steps in the sampling procedure of sign-problematic QMC
can be separated in the context of the many-fermion problem. To do this, we replace step (2), the calculation of thermodynamic 
observables, with supervised machine learning on configuration data produced in step (1). Neural networks have recently been demonstrated capable of 
discriminating between classical phases of matter, through direct training on Monte Carlo configurations \cite{Carrasquilla2016,Wang2016}.  
In this paper, we employ auxiliary-field QMC techniques to sample statistical instances of the wavefunction of a fermionic system. We 
then train a convolutional neural network (CNN) to discriminate between two fermionic phases, which are known ground states for certain 
parameters of a fermionic quantum lattice model, directly with QMC samples of the Green's function. Once trained, the CNN can provide a 
prediction, for instance, of the parametric location of the phase transition between the two phases, which we demonstrate 
for a number of Hubbard-type quantum lattice models with competing itinerant and charge-ordered phases.  
Importantly, this robust prediction of quantum critical points appears 
to work even for systems where the Monte Carlo sampling of conventional observables is plagued by a severe sign problem.
Such a machine learning approach to the QMC sampling of many-fermion systems thus allows one to determine whether crucial information about
the ground state of the many-fermion system is truly lost in the sampling procedure, or whether it can be retrieved in physical entities beyond statistical
estimators, enabling a supervised learning of phases despite the presence of the sign problem.
\\


\noindent{\bf Circumventing the Fermion-Sign Problem\\}
To begin, consider a $d$-dimensional fermionic quantum system, which can be generically written in terms of a classical 
statistical mechanics problem defined on a phase space with configurations $C$ in $d+1$ dimensions. The partition function of the 
quantum system can thereby be expressed as a sum of statistical weights over classical configurations, i.e.~$Z=\sum_C W_C$.
Unlike classical systems, for quantum Hamiltonians the weights $W_C$ can be both positive and negative (or even complex),
which invalidates the usual Monte Carlo interpretation of $W_C/Z$ as a probability distribution. 
In principle, a stochastic interpretation can be salvaged by considering a modified statistical ensemble with probability 
distribution $P_C \propto |W_C|$ and concomitantly moving the sign of $W_C$ to the observable 
\begin{eqnarray}
	\langle O \rangle & = & \frac{\sum_C O_C \cdot W_C }{\sum_C W_C} = \frac{\sum_C O_C \cdot {\rm sign}(W_C) \cdot |W_C|}{\sum_C {\rm sign}(W_C) \cdot |W_C|} \nonumber\\
				  & = & \frac{\langle {\rm sign} \cdot O\rangle_{|W|}}{\langle {\rm sign}\rangle_{|W|}} \,.
\label{expvalue}
\end{eqnarray}
This procedure, although formally exact, introduces the QMC sign problem as a manifestation of the ``small numbers problem", 
where the numerator and denominator in the last expression both approach zero exponentially in system size $N$ and inverse temperature $\beta$ ~\cite{Hirsch1982,Loh1990},
i.e.~we have
\begin{equation}
	\langle {\rm sign}\rangle_{|W|} = \exp( -\beta N \Delta f) \,,
\end{equation}
where $\Delta f$ is the difference in the free energy densities of the original fermionic system and the one with absolute weights.
Thus resolving the ratio in Eq.~\eqref{expvalue} within the statistical noise inherent to all Monte Carlo simulations becomes exponentially hard. 
The advantage of importance sampling, which often translates into polynomial scaling, is lost. 

In this work, instead of attempting to obtain exact expectation values 
of physical observables, or attempting to find a basis where the weights $W_C$ are always non-negative or that ameliorates the calculation of 
$\langle {\rm sign}\rangle_{|W|}$, we introduce a basis-dependent ``state function''  $F_C$ whose goal is to associate configurations $C$ with the most likely phase of matter they belong to for a given Hamiltonian. 
More precisely, we assume that there exists a function $F_C$ such that  
its expectation value in the modified ensemble of absolute weights
\begin{equation}
	\langle F \rangle_{|W|} =  \frac{\sum_C F_C \cdot |W_C|}{\sum_C |W_C|}
 \label{Fabs} 
\end{equation} is $1$ when the system is deep in phase $A$ and $0$ when the system is deep in the neighboring phase $B$. 
Around the critical point separating phase $A$ from $B$, $\langle F \rangle_{|W|}$ crosses over from one to zero. 
The value $\langle F \rangle_{|W|}=1/2$ indicates 
that the function can not make a distinction between phases $A$ and $B$, and therefore assigns equal probability to both phases.
We therefore interpret this value as locating the position of the transition separating the two phases in parameter space \cite{Footnote:InflectionPoint}. 
In practice, we use a deep CNN to approximate the state function $F$, which is trained on ``image'' representations of  
configurations $C$ sampled from the modified ensemble $|W_C|/\sum_C |W_C|$ in the two different phases $A$ and $B$. We explore
several choices for this image representation including color-conversions of the auxiliary field encountered in determinental Monte Carlo approaches, the Green's function as well as the Green's function multiplied by the sign.
If the above procedure indeed allows the crafting of such a state function $F$, then one has found a path to a {\it sign-problem avoiding} discrimination of the two phases and their phase transitions through the evaluation of $\langle F \rangle_{|W|}$. 
\\


\noindent{\bf Convolutional Neural Networks}\\
Artificial neural networks have for some time been identified as the key ingredient of powerful pattern recognition and machine learning algorithms \cite{Schmidhuber2015,Nielsen2016}.
Very recently, neural networks and other machine learning algorithms have been brought to the realm of statistical physics. On a conceptual level, parallels between deep 
learning and renormalization group techniques have been explored \cite{Beny2013,Mehta2014}, while on a more practical level machine learning algorithms have been applied 
to model potential energy surfaces \cite{Behler2007}, relaxation in glassy liquids \cite{Schoenholz2016} or the identification of phase transitions in classical many-body 
systems \cite{Carrasquilla2016,Wang2016}. Boltzmann machines, as well as their quantum extensions~\cite{Amin2016}, have been applied to statistical mechanics 
models~\cite{Torlai2016} and quantum systems \cite{Carleo2016}.  In addition, new supervised learning algorithms inspired 
by tensor-network representations of quantum states have been recently proposed \cite{Stoudenmire2016}. 
\begin{figure}[t]
	\includegraphics[width=\columnwidth]{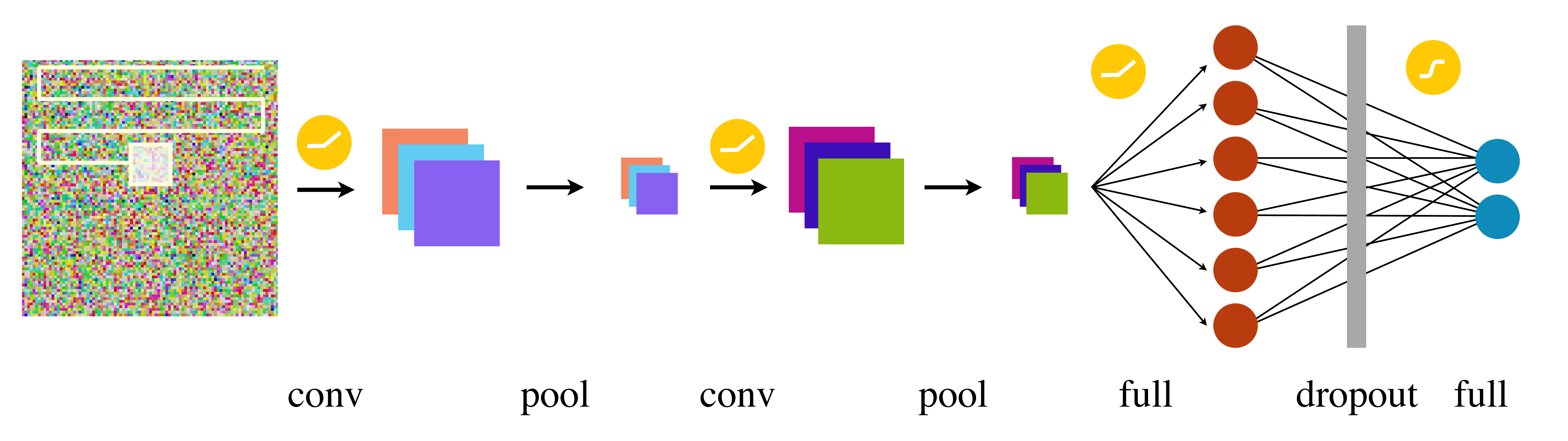}
	\caption{(Color online) Schematic illustration of the neural network used in this work. 
			A combination of convolutional (conv) and max pooling layers (pool) is first used to study the image, 
			before the data is further analyzed by two fully connected neural networks separated by a dropout layer. 
			The convolutional and the first fully connected layer are activated using rectified linear functions, 
			while the final layer is activated by a softmax function.
			\label{fig:cnn_architecture}}
\end{figure}

In machine learning, the goal of artificial neural networks is to learn to recognize patterns in a (typically high dimensional) data set.  CNNs, in particular, are 
nonlinear functions which are optimized (in an initial ``training" step) such that the resulting function 
$F$ allows for the extraction of patterns (or ``features'') present in the data. Here we take this approach to construct a function $F$, represented as 
a deep CNN, that  allows the classification of many-fermion phases as outlined in the previous section. Our choice of employing a deep CNN is rooted in the above observation that the configurations generated from a quantum Monte Carlo algorithm can be often interpreted as ``images''.
As we explain below in more 
detail, our analysis can be regarded as an image classification problem -- an extremely successful application of CNNs.

The architecture of the CNN we use is depicted schematically in Fig.~\ref{fig:cnn_architecture} with a more detailed technical discussion of the individual components
presented in the Methods section. We feed the CNN with Monte Carlo configurations (illustrated on the left), which, processed through the network, provide a two-component 
softmax output layer (on the right). The two components of this function, which by construction always add up to one, can be interpreted as the probabilities 
that a given configuration belongs to the two different phases under consideration and can thus be used for classification. In the initial training step, we 
optimize the CNN on a set of (typically) $2 \times 8192$ representative configurations sampled deep in the two fermionic phases. 
The question of which fundamental features, contained in the Monte Carlo configurations, are used in the resulting function $F$ to characterize the phases under consideration, is automatically discovered during the training 
procedure (and beyond our direct influence).\\


\noindent{\bf Machine learning fermionic quantum phases\\}
We apply this QMC + machine learning framework to a family of Hubbard-like fermion models where the competition between kinetic and potential 
terms gives rise to a phase transition between an itinerant metallic phase and a charge-ordered Mott insulating phase. As a first example
we consider a system of spinful fermions on the honeycomb lattice subject to the Hamiltonian
\begin{equation}
	H =  K + V = -t\sum\limits_{\langle i, j\rangle, \sigma} c_{i,\sigma}^\dagger c_{j,\sigma}^{\phantom \dagger} + U \sum\limits_{i} n_{\uparrow, i} n_{\downarrow, i} \,,
	\label{eq:spinfulHamiltonian}
\end{equation}
with a kinetic term $K$ and a potential term $V$.
At zero temperature and half-filling, this system is well known to undergo a quantum phase transition from a Dirac semi-metal to an insulator with antiferromagnetic 
spin-density wave (SDW) order at $U_c/t \approx 3.85$ ~\cite{Sorella2015}. For convenience, we will set $t=1$ in the following.

To sample configurations for different values of the tuning parameter $U$ we employ determinantal quantum Monte Carlo (DQMC)
in its projective zero-temperature formulation. 
In this scheme, a carefully chosen test wave function $\ket{\psi_T}$ is projected onto the actual ground state function $\ket{\psi}$
\begin{equation}
	\ket{\psi} = e^{-\theta H}\ket{\psi_T} \,.
\end{equation}
To compute this projection, we first apply a Trotter decomposition to discretize the projection time $\theta$ into $N_\tau = \theta / \Delta\tau$ 
time steps and separate the kinetic and potential terms
\begin{equation}
e^{-\theta H} = \prod\limits_{n = 1}^{N_\tau} e^{-\Delta\tau K} e^{-\Delta\tau V} \equiv B(\theta) \,.
\end{equation}
The quartic interaction term is then decomposed by applying a Hubbard-Stratonovich (HS) transformation on each on-site interaction $V_i$ and on each time slice $\tau$
\begin{equation}
e^{-\Delta\tau V_i} = \dfrac{1}{2}\sum\limits_{s = \pm 1}\prod\limits_{\sigma = \uparrow,\downarrow} e^{-V_i(s, \tau, \sigma)} \label{eq:hs_sample},
\end{equation}
introducing one auxiliary variable $s= \pm 1$ per site and separating the two spin species $\sigma$. 
The entirety of the auxiliary variables makes up the Hubbard Stratonovich field and will be denoted as $\mathbf{s}$ in the following. 
The probability for choosing a configuration is given by
\begin{equation}
p(\mathbf{s}, \mathbf{s^\prime}) = \dfrac{\braket{\psi(\mathbf{s})}{\psi(\mathbf{s^\prime})}}{\sum\limits_{\mathbf{s}\mathbf{s^\prime}}{\braket{\psi(\mathbf{s})}{\psi(\mathbf{s^\prime})}}},
\end{equation}
where $\mathbf{s}$ and $\mathbf{s}^\prime$ denote the Hubbard-Stratonovich fields associated with the projection of the wavefunction used as bra and ket, respectively.
The weight of the configuration $\braket{\psi(\mathbf{s})}{\psi(\mathbf{s^\prime})}$ evaluates to a determinant 
\begin{equation}
\braket{\psi(\mathbf{s})}{\psi(\mathbf{s^\prime})} = \det{\left[P^\dagger B(\theta, \mathbf{s^\prime})B(\theta, \mathbf{s^\prime})P\right]}
\label{eq:weight}
\end{equation}
where $P$ is the matrix representation of the test wave function $\ket{\psi_T}$. 
For auxiliary field approaches the modified statistical ensemble of absolute weights implies that the sign
of the fermionic determinant will be ignored -- importantly, such a modified ensemble {\it retains} the fermionic exchange statistics, but becomes insensitive to the parity of the total number of fermionic exchanges for a given configuration (which is precisely what is reflected in the sign of the determinant). This should be contrasted to world-line QMC approaches where the modified ensemble weighted by $|W_C|$ would not preserve any fermionic exchange statistics at all, but effectively sample a {\it bosonic} system.

\begin{figure}
	\includegraphics[width=\columnwidth]{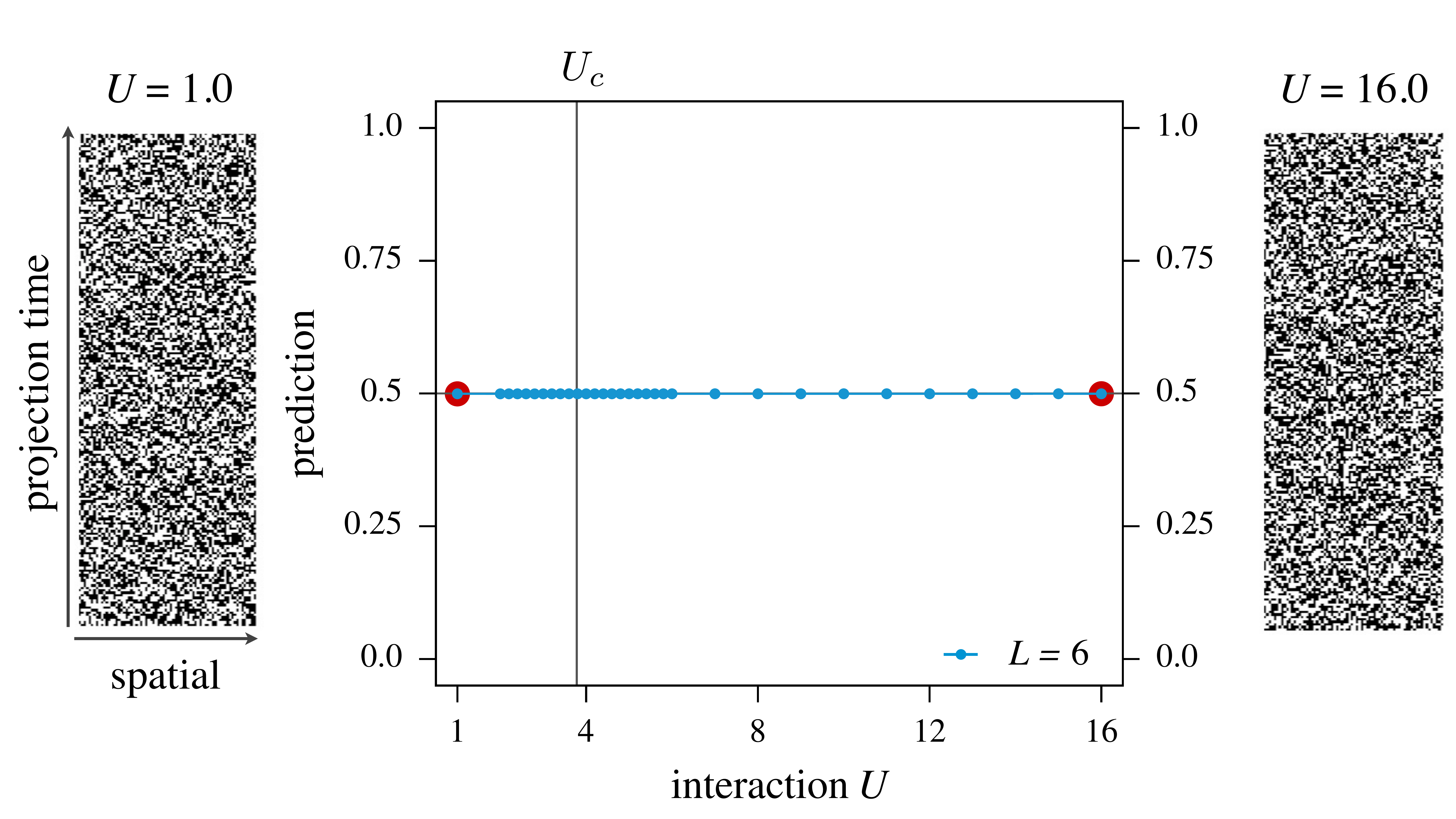}
	\caption{(Color online) 
			Results from training the neural network on Hubbard-Stratonovich field configurations of a spinful Hubbard model on a 
			$2\cdot 6\times 6$ lattice with on-site interaction $U$.
			Reference points for training were $U=1.0$ and $U=16.0$, marked by red dots in the figure. 
			Despite intensive training, the network depicted in Fig.~\ref{fig:cnn_architecture} is unable to distinguish the auxiliary field 
			configurations of the two reference points and as a consequence can not be used to discriminate between the two phases. 
	\label{fig:spinful_cnn_aux}}
\end{figure}

In order to implement our machine learning approach, we begin by choosing 
the classical configuration space $C$ 
over which the expectation values in Eqs.~\eqref{expvalue} and \eqref{Fabs} are taken.
An obvious candidate is the auxiliary field ${\bf s}$.
This approach is illustrated in Fig.~\ref{fig:spinful_cnn_aux}, where the CNN has been trained 
at parameters $U = 1$ and $U=16$, i.e.~deep within the Dirac semi-metal and 
the antiferromagnetic SDW phase, respectively. 
The side panels show representative reference configurations of the auxiliary field at each of these training parameters.
Interestingly, the configurations displayed in Fig.~\ref{fig:spinful_cnn_aux} show no discernible difference between the two 
auxiliary field configurations, apparent to the human eye. Indeed, we find that optimizing the CNN of Fig.~\ref{fig:cnn_architecture} to extract information directly from these auxiliary field configurations does not yield a function $F$ that allows one to distinguish between the two phases. 
This apparent inability is possibly rooted in the particular choice of the employed Hubbard-Stratonovich transformation, which preserves SU(2) spin symmetry by decoupling in the charge channel. There is a multitude of alternative Hubbard-Stratonovich transformations that one could choose for this problem that would result in different configurations of the auxiliary field.
It is thus possible, yet not guaranteed, that the training could succeed for another choice of the Hubbard-Stratonovich transformation
\cite{Footnote:HS}.
\begin{figure}
	\includegraphics[width=\columnwidth]{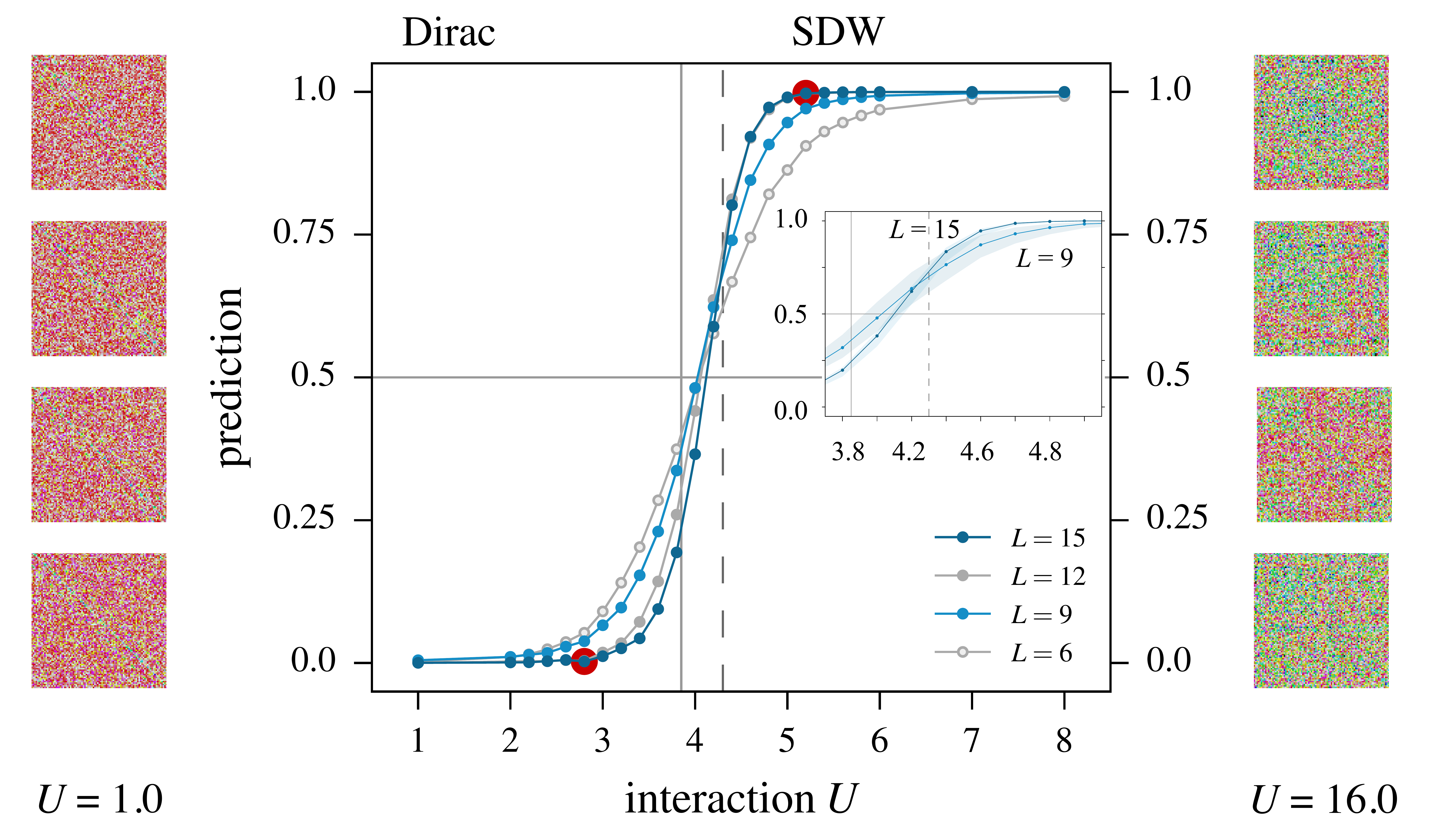}
	\caption{(Color online) 
			Machine learning of the phase transition from a semi-metal to an antiferromagnetic insulator in the spinful Hubbard model \eqref{eq:spinfulHamiltonian} on a honeycomb lattice
			using the Green's function approach (see main text). 
			Visualized in the side panels are representative samples of the Green's function (calculated from the auxiliary field)
			 for a $2\cdot 9 \times 9$ system in the two respective phases. The complex entries of 
			these matrices are color-converted by interpreting their absolute value as the  hue of the color while their angle is chosen as the saturation 
			(HSV coloring scheme \cite{Foley:1990:CGP:83821}). 
			The main panel shows the output of the discriminating function $F$ obtained from a CNN trained for parameters in the two fermionic phases (indicated by the red dots).
			Data for different system sizes $2\cdot L \times L$ are shown where the colors were selected to highlight an apparent even-odd effect in the linear system size.
			The vertical solid line indicates the position of the phase transition in the thermodynamic limit~\cite{Sorella2015}, 
			while the dashed line marks the position at which the antiferromagnetic order breaks down ~\cite{Assaad2012}
			for the finite system sizes of the current study.
	\label{fig:spinful_cnn_greens}}
\end{figure}
%


To alleviate this difficulty, we instead consider the Green's function $G(i, j) = \langle c_i^{\phantom\dagger}c_j^\dagger \rangle$ as input for 
our machine learning approach. The Green's function is an essential quantity in statistical physics, which allows e.g.~for the calculation of equal-time correlation functions,
and while it can easily be calculated from a given auxiliary field configuration it is not sensitive to the specifics of the Hubbard-Stratonovich transformation.
Instead of the bare auxiliary fields, we thus train the CNN on the unprocessed complex valued Green's matrices $G_s(i, j) = \langle c_i^{\phantom\dagger}c_j^\dagger \rangle_s$
calculated for a given auxiliary field configuration $s$. For the training, we used $2\times 8192$ ($2\times 4096$ for $L=15$) samples of the Green's function.
This modified approach gives a striking improvement in results, as illustrated in Fig.~\ref{fig:spinful_cnn_greens}.
The side panels now show representative examples of the Green's matrices $G_s(i, j)$ for the two coupling parameters well inside the two respective fermionic phases.
For the purpose of visualization, we convert the complex-valued entries of the Green's matrices to a polar representation which are then interpreted as HSV colors and finally 
converted to RGB for illustration~\cite{Foley:1990:CGP:83821}. 
Contrary to the visual inspection of the auxiliary field configuration in Fig.~\ref{fig:spinful_cnn_aux}, the image-converted Green's function exhibits a clearly visible distinction for the two phases.
Indeed the CNN trained and applied to the image-converted Green's function now succeeds in discriminating the two phases by producing a function $F$ that indicates a phase transition around a value of the interaction $U\approx 4.1 \pm 0.1$. For a given finite system size $L$, we identify the location of the phase transition with the parameter $U$ for which the averaged state function $F$ is $1/2$, i.e.~the parameter for which the CNN cannot make any distinction between the two phases and therefore assigns equal probability to both phases. These estimates for the location of the phase transition and their finite-size trends are in good agreement with the critical value of $U_c(L=15) \approx 4.3$ obtained from Monte Carlo simulations for similar system sizes \cite{Assaad2012} and slightly above the critical value $U_c(L\to\infty) \approx 3.85$ of the thermodynamic limit \cite{Sorella2015}.\\


%
\begin{figure}
	\includegraphics[width=\columnwidth]{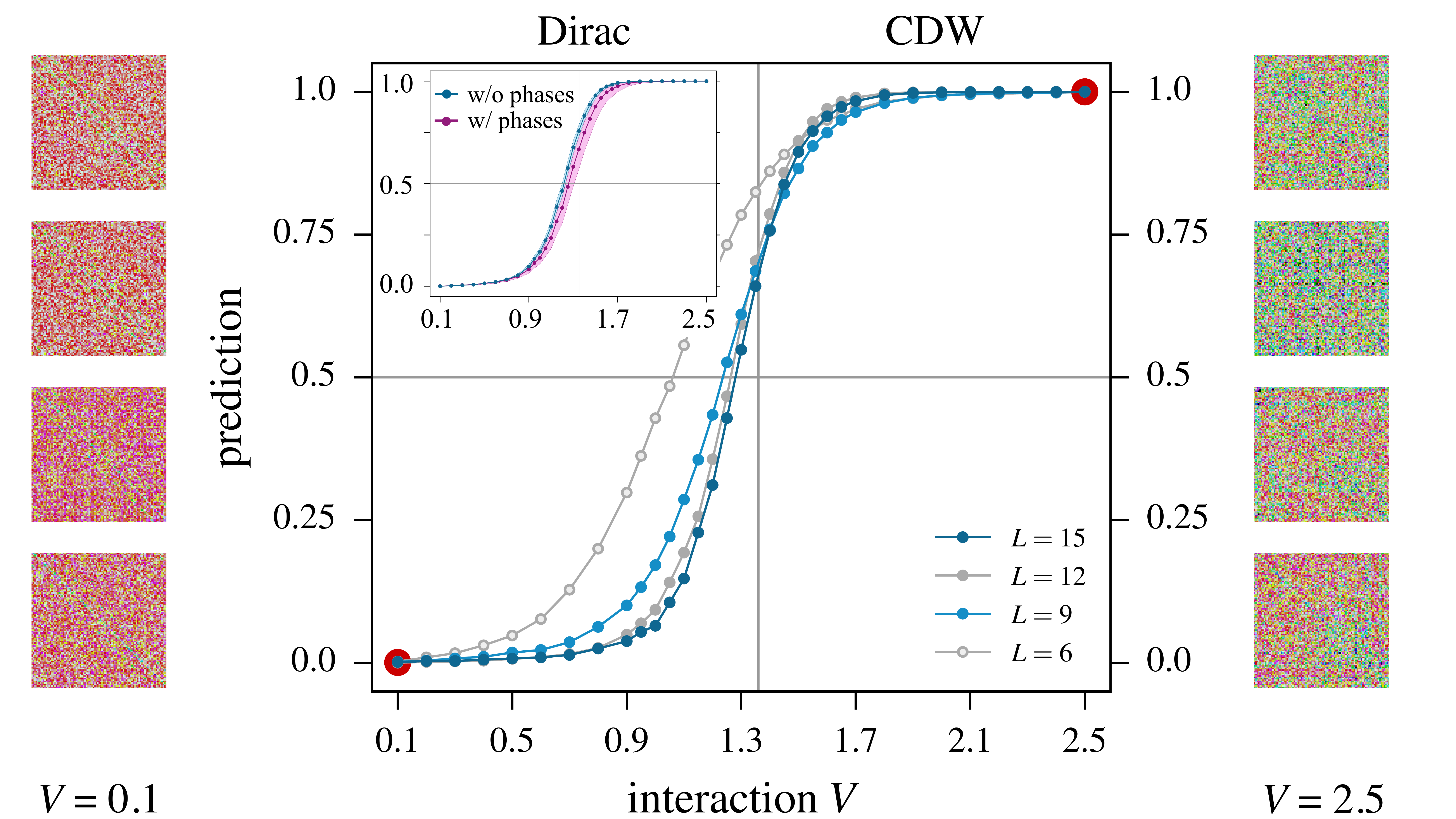}
	\caption{(Color online) 
		Prediction of a CNN for the phase transition from a Dirac semi-metal to a charge density wave (CDW) ordered state 
		in the half-filled spinless fermion Hubbard model \eqref{eq:spinlessHamiltonian} on the honeycomb lattice of size \
		$2\cdot L\times L$. 
		The CNN has been trained on 8192 representative samples of the bare Green's function deep inside the two phases
		(indicated by the red dots). The images in the left and right columns are color-converted instances of the Green's function
		used in the training.
		The inset shows a comparison of the prediction for the $L=9$ system when feeding the CNN with the bare Green's function
		or the Green's function multiplied by the relative sign / complex phase associated with each configuration (of a given Markov chain). 
	\label{fig:spinless_cnn_greens}}
\end{figure}

\noindent{\bf Sign-problematic many-fermion systems\\}
We now turn to many-fermion systems that exhibit a sign problem in the conventional QMC + statistical analysis approach,
and ask to what extent the QMC + machine learning framework is sensitive to this sign problem.
Simple example systems 
of this sort are {\it spinless} fermion models, which typically exhibit a severe sign problem in the conventional complex fermion basis.
We first consider a half-filled honeycomb system subject to the Hamiltonian,
\begin{equation}
	H =  -t\sum\limits_{\langle i, j\rangle} c_i^\dagger c^{\phantom\dagger}_j+ V \sum\limits_{\langle i, j\rangle} n_i n_j.
	\label{eq:spinlessHamiltonian}
\end{equation}
The competition between the kinetic term (which we again set to $t=1$) and 
a repulsive nearest neighbor interaction $V$ drives the system through a quantum phase transition \cite{GrossNeveu} 
separating a semi-metallic state for $V < V_c$ from a charge density wave (CDW) state for $V > V_c$.
Interestingly, this model can be made to be sign-problem free through a basis transformation
to a Majorana fermion basis \cite{Wang2014}, which allows for a precise estimation of the critical repulsion $V_c \approx 1.36$
directly from QMC observables \cite{Wang2014,Li2015,Motruk2015,Capponi2015,Broecker2016}.
For the purpose of this paper, we will not perform this transformation, but rather sample the model in its sign-problematic formulation in
the conventional complex fermion basis.

Analogous to our procedure for the sign-problem free case of spinful fermions, we first train the CNN on representative samples of the Green's function for parameters deep within the two phases.
In the below, we generate 8192 (4096 for $L=15$) samples for $V = 0.1$ (semi-metal) and $V = 2.5$ (CDW) from DQMC simulations using the modified statistical ensemble of absolute weights $|W_C|$. 
Thus optimized, we then feed unlabeled configurations from several different interaction values $0.1 < V < 2.5$ 
and ask the neural network to predict to which phase a particular configuration belongs.

At this point, a decision has to be made about how to provide information about the sign of each configuration to the CNN.
We explore two options.  First, we multiply each Green's matrix $G_s(i, j)$ with the sign (in general a complex phase) associated with the underlying configuration, i.e. ${\rm sign}(W_C)$, for a given Markov chain.  
Second, we ignore the sign altogether, and feed the CNN the ``bare'' Green's function without any information about the sign.
Surprisingly, as illustrated in the inset of Fig.~\ref{fig:spinless_cnn_greens}, the state function $F$ for the phase-multiplied Green's functions does not exhibit a notable improvement in predicting the position of the phase transition over the bare Green's function.  
While the function moves slightly in parameter space, it also acquires a much broader spread (estimated from averaging over 12 epochs, see the Methods section) \cite{Footnote:RandomPhases}. 
Considering the data for different system sizes in Fig.~\ref{fig:spinless_cnn_greens} one can determine a quantitative estimate of the location of the fermionic phase transition, which is in very good agreement with the Monte Carlo results.  
This convincingly demonstrates that the CNN is capable of providing a high-quality state function $F$ discriminating the two fermionic phases,
even when the sign content of the configurations is ignored.
Importantly, we note that the approach with bare Green's matrices can provide a significant 
gain in computational efficiency over that which includes information about the relative sign of individual configurations,
by sampling multiple parallel Markov chains. 
Thus, in light of the results of Fig.~\ref{fig:spinless_cnn_greens} (inset), which show 
no systematic improvement of the state function $F$ given additional information on the sign structure,
we choose to show results for the bare Green's functions in the examples below.
The fact that such an approach produces a highly accurate state function $F$ is 
a striking demonstration of the power of QMC + machine learning, even in models afflicted with a serious sign problem.

\begin{figure}
	\includegraphics[width=\columnwidth]{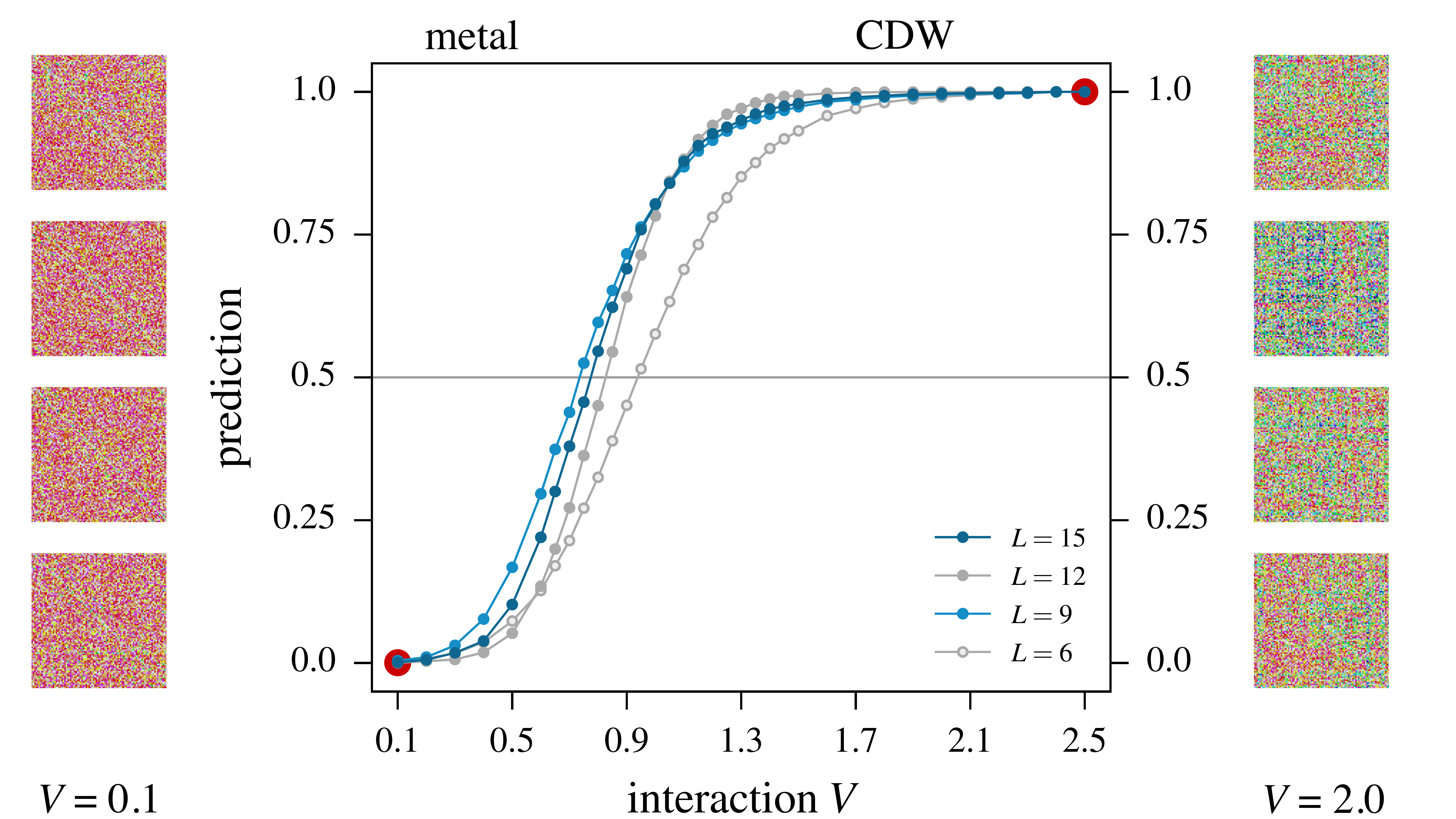}
	\caption{(Color online) 
		CNN-based identification of the phase transition in the one-third filled, spinless Hubbard model \eqref{eq:spinlessHamiltonian} on the honeycomb lattice with nearest-neighbor repulsion $V$. 
		The side panels show representative samples of the Green's functions at the two reference points 
		$V = 0.1$ and $V = 2.5$. 
		The network finds two clearly separated phases that extend of which we know the weak coupling 
		phase to be metallic. 
		\label{fig:spinless_cnn_greens2}}
\end{figure}

Next, we consider the spinless fermion system of Eq.~\eqref{eq:spinlessHamiltonian} at one-third filling. Going below half-filling turns the itinerant phase for small coupling $V$ into a conventional metal with a nodal Fermi line, while for large $V$ we still expect some sort of CDW-ordered Mott insulating state. In contrast to half-filling, the one-third-filled system has no known sign-free (Majorana) basis. Applying our QMC + machine learning approach to this problem, we again find that a state discriminating function $F$ can be identified by a properly optimized CNN.
This procedure indicates the existence of a phase transition around $V_c \approx 0.7 \pm 0.1$ as illustrated in Fig.~\ref{fig:spinless_cnn_greens2}, which matches a recent estimate from entanglement calculations \cite{Broecker2016}.
The precise nature of the Mott insulating phase at large $V$ has so far remained elusive, which unfortunately is not altered by
the supervised learning approach employed in the current study. 

\begin{figure}
	\includegraphics[width=\columnwidth]{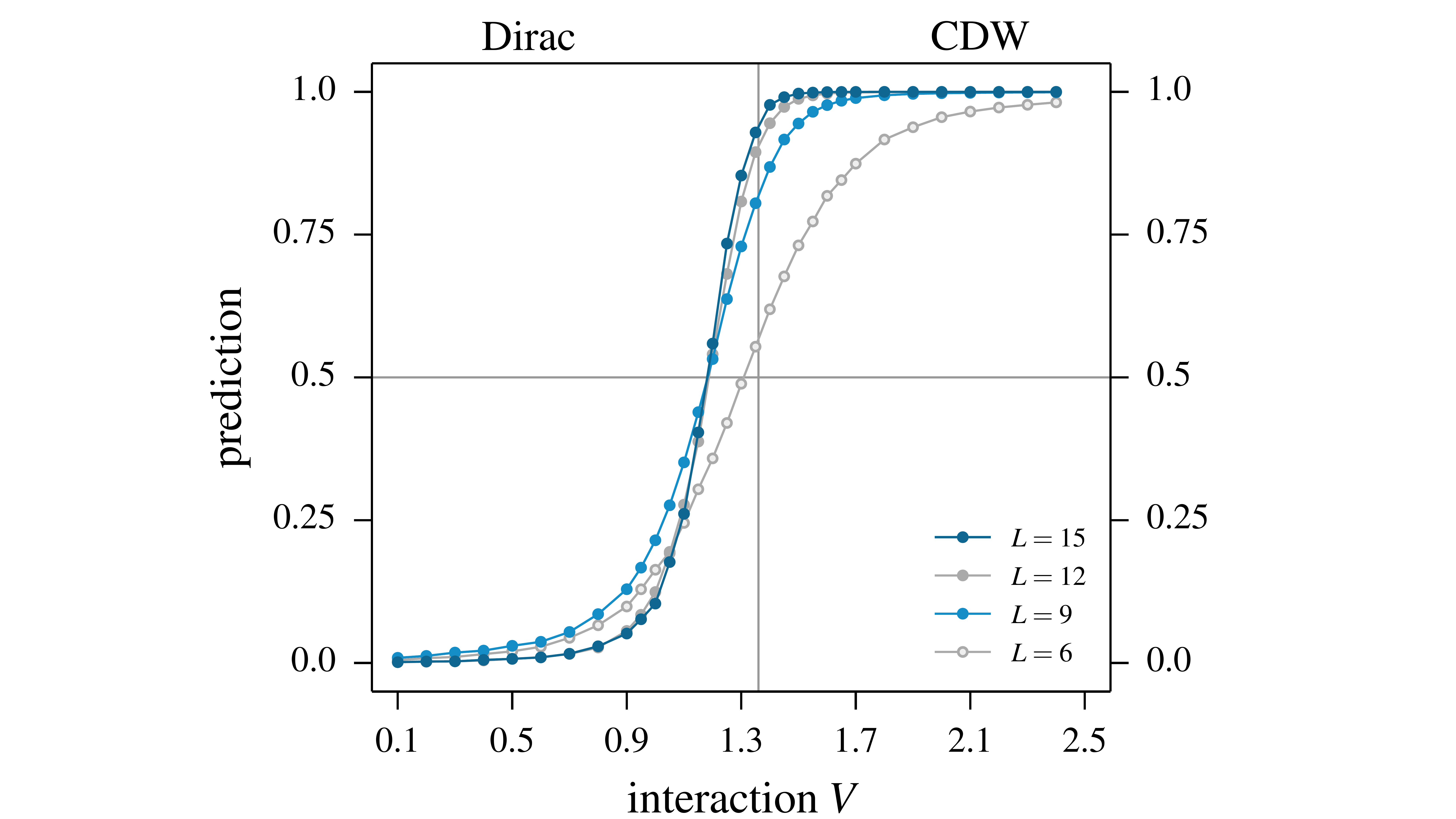}
	\caption{(Color online) 
		An example of transfer learning in an artificial neural network. 
		A CNN that was trained to discriminate the phases of the sign-problem free, 
		{\it spinful} Hubbard model \eqref{eq:spinfulHamiltonian} and then applied to 
		identify the phases and phase transition of the sign-problematic, {\it spinless} 
		Hubbard  model \eqref{eq:spinlessHamiltonian}. 
		The network is found to reliably distinguish the fermionic phases of the spinless 
		model and provides a relatively accurate estimate for  the location of the phase 
		transition (the vertical line indicates the location of the transition in the 
		thermodynamic limit of infinite system size). 		
	\label{fig:spinless_cnn_cross_check}}
\end{figure}

Finally, we explore whether we can generate ``transfer learning" by training a neural network on one model, then 
using the trained network to discriminate phases from configurations produced for an entirely different Hamiltonian.
This approach was highly successful for neural networks trained with classical Ising configurations in Ref.~\cite{Carrasquilla2016}.
Here, using samples of the Green's function, we train a CNN to discriminate the fermionic phases of the sign-problem free, spinful fermion model \eqref{eq:spinfulHamiltonian} and then apply the trained network for supervised learning on the sign-problematic, spinless fermion model \eqref{eq:spinlessHamiltonian}.
This procedure seems justified based on the fact that
at half-filling the two models exhibit similar physics, with the potential energy driving a Gross-Neveu type phase transition from a Dirac semi-metal to a SDW/CDW charge-ordered phase, respectively. 
Results for the predictions of the averaged state function are illustrated in Fig.~\ref{fig:spinless_cnn_cross_check},
which shows that the CNN is capable of reliably distinguishing the fermionic phases of the spinless model,
even producing a rough estimate for the location of the phase transition.
Thus, we find that this approach indeed allows for a certain level of transfer learning
between sign-problem free and sign-problematic Hamiltonians,
suggesting a fruitful area of future study on the relationship between supervised machine learning and the sign problem.\\

%


\noindent{\bf Discussion\\}
We have introduced a powerful numerical scheme to reliably distinguish fermionic phases of matter by a combination of quantum Monte Carlo sampling 
and a subsequent machine learning analysis of the sampled Green's functions via a convolutional neural network. Our numerical experiments for a family
of Hubbard-type models demonstrate that this approach extends to sign-problematic many-fermion models that are not amenable to the conventional QMC
approach of sampling and statistical analysis. 
These findings thereby provide a perspective on the information content of the sampled {\it ensemble} of Green's functions. In contrast to a conventional 
statistical physics analysis, in which equal-time correlation functions calculated from this ensemble of Green's functions exhibit a statistical uncertainty 
so large that they are rendered completely unusable, the machine learning approach demonstrates that the same ensemble of Green's functions 
holds sufficient information to positively discriminate fermionic phases.
This Green's function based machine learning approach is very general and can be applied to QMC flavors beyond the auxiliary field techniques applied
in the current work. In particular, this approach can be readily adapted by world-line Monte Carlo approaches which 
are highly successful in the
study of quantum magnets and bosonic systems. 
We expect that this QMC + machine learning approach will establish itself as a robust tool for quickly and semi-automatically mapping out phase diagrams 
of quantum many-body systems, and in the future will become a key ingredient of our numerical toolbox complementing existing statistical physics approaches.\\


\begin{acknowledgments}
\noindent {\bf Acknowledgements}\\
P.B. acknowledges partial support from the Deutsche Telekom Stiftung 
and the Bonn-Cologne Graduate School of Physics and Astronomy (BCGS).
The Cologne group was partially supported by the DFG within the CRC network TR 183 (project B01).
The numerical simulations were performed on the CHEOPS cluster at RRZK Cologne.
R.M. acknowledges support from NSERC and the Canada Research Chair program.
Additional support was provided by the Perimeter Institute for Theoretical Physics. 
Research at Perimeter Institute is supported by the Government of Canada through the Department of Innovation, Science and Economic Development Canada and by the Province of Ontario through the Ministry of Research, Innovation and Science.
R.M. and S.T. thank the Nordic Institute for Theoretical Physics (NORDITA) for hospitality during the workshop ``From Quantum Field Theories to Numerical Methods" where  this project was invigorated in its early stages.\\
\end{acknowledgments}

\noindent {\bf Author contributions\\} 
P.B. developed the ideas underlying this manuscript in discussion with all authors. P.B. and J.C. implemented all simulations codes, P.B. collected the numerical data. 
Data analysis and interpretation was jointly done by all authors. All authors contributed to the writing of the manuscript.\\


\noindent {\bf Methods\\} 
\noindent{\it Machine learning. }
Neural networks come in a huge variety of different architectures; 
precisely which setup to choose for a specific problem is answered by selecting the empirically most successful architecture. 
In this paper, we started with a setup, see Fig.~\ref{fig:cnn_architecture}, that is successfully used to classify images such as the CIFAR-10 dataset~\cite{cifar10}.  
Its network architecture consists of two main components -- a convolutional and a fully connected part.
The convolutional part processes the data by a combination of two convolutional 
and max pooling units. Both of these units are activated by a rectified linear function (relu) and have filters of size $3\times 3$.  The total number of filters 
is $32$ for the first and $64$ for the second. 
The data is then fed into a fully connected, relu activated layer of $512$ neurons. 
To avoid overfitting, we applied a dropout regularization at a rate of $0.5$ to this layer. 
At the output of the CNN we consider a fully connected softmax layer. 
The optimization of the neural network is performed using a cross entropy as a cost function and ADAM~\cite{Kingma2014} as a particularly efficient variant of the stochastic gradient at a learning rate of $\gamma = 0.0001$.
The network was trained over $16$ epochs and results were averaged over the last 8 epochs.
Our numerical implementation of the neural network is based on the TensorFlow library \cite{TensorFlow}.\\


\noindent{\it Determinantal Quantum Monte Carlo. }
For our DQMC simulation, we use a projective algorithm with a discretization step of $\Delta\tau = 0.1$ and a projection time $\theta = 10$. 
Thus, the auxiliary field for the spinful Hubbard model is of size $2\cdot L^2 \times 200$. The Green's functions are of size $2\cdot L^2 \times 2\cdot L^2$.
The test wave function $\ket{\psi_T}$ is generated by taking the quadratic part of the Hamiltonian and randomizing the hopping strengths strongly enough such that the eigenvalues of adjacent states are separated by more than $10^{-3}$. 
The eigenvectors corresponding to the lowest $N_{\text{particles}}$ eigenvalues are used for the test wave function.

The Hubbard-Stratonovich transformation is applied to each quartic operator, introducing the auxiliary field. 
For the models studied in this paper, one such transformation has to be carried out for each site or each bond, respectively, and on each slice in projection time. 
There are various ways to perform this transformation, in particular, one is often free to choose the channel one performs the transformation in and what type of field should be created. 
One possible realization is to decouple a density-density interaction of strength $U$ with general indices $\alpha$ and $\beta$ denoting for example spin and / or lattice site in the following way
\begin{equation}
e^{-\Delta\tau U n_\alpha n_\beta} = \dfrac{1}{2}\sum\limits_{s = \pm 1}\prod\limits_{a = \alpha,\beta} e^{-\left(s\lambda  + \frac{U\Delta\tau}{2}\right)\left(n_a - \frac{1}{2}\right)}.\label{eq:hs_sample}
\end{equation}
where the auxiliary variable $s$ is in $\{\pm 1\}$ and $\lambda$ is a constant related to $U$.
This transformation results in complex weights for $U > 0$, i.e. a repulsive interaction. 
In the spinful Hubbard model at half filling, the product of the phases of the two determinants and the prefactor result in an overall prefactor of $1$ for the weight, i.e. there is no sign problem. 
This changes drastically once one moves away from half filling or takes away one of the fermion species, resulting in a severe sign or phase problem. 
An alternative transformation that allows us to work with real numbers only works by decoupling in the magnetization channel. 
While at first look computationally favorable (because of the real numbers), it turns out that the convergence of magnetic observables is significantly better in the complex case, 
as it retains the $SU(2)$ symmetry explicitly while in the real case this symmetry is only restored after the summation over all configurations has been carried out.

For the phase sensitive calculations, one can in principle calculate the absolute phase of a weight from the determinant in Eq.~\eqref{eq:weight}. However, this approach is found to be plagued by numerical instabilities making its computation prohibitively
expensive in terms of computing resources.
Alternatively, one may track the changes in the phase along the Markov chain and thus calculate the {\it relative} phase with respect to an initial phase for each  configuration visited in the Markov chain. 
The change in phase $\phi^\prime / \phi$ is given by the phase of the ratio of weights $W(C^\prime) / W_C$ between the current configuration $C$ and a proposed configuration $C^\prime$.
Using this quantity, the initial phase $\phi$ is updated by multiplying $\phi$ with the ratio of phases for adjacent steps on the Markov chain
\begin{equation}
\phi \xrightarrow{\cdot\frac{\phi^\prime}{\phi}} \phi^\prime\xrightarrow{\cdot\frac{\phi^{\prime\prime}}{\phi^\prime}}\dots\xrightarrow{\cdot\frac{\phi^{n + 1}}{\phi^{n}}}\phi^{n + 1} \,.
\end{equation}
Using the relative phase has the advantage that it is possible to compute this quantity with very high accuracy, while it is not expected to change any of the physics (a global transformation of the phase of the weights is compensated when normalizing the partition or wave function).


\bibliography{machine_learning}

\begin{thebibliography}{40}
\bibitem [{Foo({\natexlab{a}})}]{Footnote:InflectionPoint}%
  \BibitemOpen
  \bibinfo {note} {{An alternative choice to identify the location of the phase
  transition for a finite system size is the parametric location of the
  inflection point of the prediction function $F$. At the inflection point the
  derivative is maximal, i.e. moving the coupling parameters slightly towards
  any one of the two phases leads to the largest possible change in the
  prediction. Such an indicator of the phase transition is commonly used e.g.
  when considering susceptibility measurements of conventional order
  parameters. For our setup we find almost no distinguishable difference in the
  location and finite-size scaling of the inflection point and the point at
  which the prediction function $F=1/2$.}}\BibitemShut {Stop}%
\bibitem [{Foo({\natexlab{b}})}]{Footnote:HS}%
  \BibitemOpen
  \bibinfo {note} {{It is also possible that further refinement of the CNN such
  as a different set of hyper parameters, or a different convolution, may yield
  a useful $F$ for this choice of field, although we have not been successful
  in finding any.}}\BibitemShut {Stop}%
\end{thebibliography}

\end{document}